\documentclass[conference]{IEEEtran}
\IEEEoverridecommandlockouts
\usepackage{orcidlink}
\usepackage{cite}
\usepackage{amsmath,amssymb,amsfonts}
\usepackage{algorithmic}
\usepackage{graphicx}
\usepackage{textcomp}
\usepackage{xcolor}
\def\BibTeX{{\rm B\kern-.05em{\sc i\kern-.025em b}\kern-.08em
    T\kern-.1667em\lower.7ex\hbox{E}\kern-.125emX}}

\begin{document}

\title{Is the Computing Continuum Already Here?}

\author{
\IEEEauthorblockN{Jacopo Marino}
\IEEEauthorblockA{\textit{Dept. of Control and Computer Engineering} \\
\textit{Politecnico di Torino}\\
Torino, Italy \\
jacopo.marino@polito.it}

\and 

\IEEEauthorblockN{Fulvio Risso}
\IEEEauthorblockA{\textit{Dept. of Control and Computer Engineering} \\
\textit{Politecnico di Torino}\\
Torino, Italy \\
fulvio.risso@polito.it}
}

\maketitle

\begin{abstract}
The computing continuum, a novel paradigm that extends beyond the current silos of cloud and edge computing, can enable the seamless and dynamic deployment of applications across diverse infrastructures.
By utilizing the cloud-native features and scalability of Kubernetes, this concept promotes \textit{deployment} transparency, \textit{communication} transparency, and \textit{resource availability} transparency.
Key features of this paradigm include intent-driven policies, a decentralized architecture, multi-ownership, and a fluid topology.
Integral to the computing continuum are the building blocks of dynamic discovery and peering, hierarchical resource continuum, resource and service reflection, network continuum, and storage and data continuum.
The implementation of these principles allows organizations to foster an efficient, dynamic, and seamless computing environment, thereby facilitating the deployment of complex distributed applications across varying infrastructures.
\end{abstract}

\begin{IEEEkeywords}
cloud computing, computing continuum
\end{IEEEkeywords}

\section{Introduction}

Despite the huge amount of research aiming at the creation of the computing continuum (just to limit our view to relevant ongoing European projects \cite{aerOS, icos, nebulous, nemo, nephele}), apparently this concept is already here.
Indeed, we have already many applications operating seamlessly across a wide spectrum of cloud and edge infrastructures, not to mention user devices running a moltitude of applications, offering the expected service and appearing to interact without perceivable constraints.
However, this paper argues that the present reality does not fully align with our final expectations and desired outcomes for the computing continuum, mainly due to a lack of \textit{transparency}.
In fact, the current implementation of the computing continuum requires distinct variants and/or configurations for each running services, which take into account the actual location of each component.

Indeed, despite the development of universal interfaces for application orchestration, existing industry practices perceives each infrastructure (i.e., datacenter clusters, but also user devices) as isolated silos, resulting in a fragmented perception of available resources \cite{6,7}.
This fragmentation hampers the seamless deployment of fully distributed applications due to various influencing factors such as resiliency, performance considerations, latency issues over Wide-Area Networks (WANs) \cite{8,9,10}, hybrid-cloud and multi-cloud approaches \cite{13}, and non-technical factors like legal regulations and physical isolation policies.
This lack of integration significantly limits workload dynamism and inhibits the deployment of complex applications with specific requirements \cite{14,15,16}.
In the prevalent use of Kubernetes as the orchestration platform, users are burdened with the placement of pods and services, and must deal with different interfaces based on whether the service is local to the cluster or hosted externally and exposed via a public endpoint.
This introduces complexity and inconsistency in the approach, as users are required to be aware of the endpoints of the services they wish to use, thereby involving them in the intricacies of the infrastructure.
For a more seamless and user-friendly experience, an abstraction layer should be implemented to facilitate the deployment of microservices across the continuum with the same ease as operating within a Kubernetes cluster.

We argue that the computing continuum should include the ideas presented in \cite{82}, in which clusters are extended to create a (borderless) virtual environment that overcomes the presented issues, the so called \textit{liquid computing}.
The key principles of the envisioned computing continuum include \textit{deployment} transparency, \textit{communication} transparency, and \textit{resource availability} transparency.
To implement it, we acknowledge Kubernetes as the reference technology due to its cloud-native features and scalability.
By embracing this vision, organizations can achieve a more efficient and dynamic environment, standardizing the communications between services and avoiding different deployments for different endpoints.

\section{The liquid computing pillars}
Liquid computing builds upon the principles of cloud and edge computing, transcending cluster boundaries to offer a flexible computing environment.
Compared to the current computing continuum, this approach provides \textit{deployment} transparency, \textit{communication} transparency, and \textit{resource availability} transparency, thus promoting optimal utilization of available resources.

In terms of \textit{deployment} transparency, liquid computing presents an enhanced strategy for deploying multi-microservice applications.
Unlike traditional configurations that restrict subsequent location modifications without initiating a new deployment phase, liquid computing allows microservices to start in the most suitable location based on service requirements and infrastructure status.
This dynamic nature simplifies user operations, with the system autonomously determining the optimal service location.
The first building block supporting this approach is dynamic discovery and peering, which promotes decentralized governance and facilitates resource and service consumption relationships between clusters. This flexible, decentralized system enables dynamic resource allocation and usage, eliminating the need for manual coordination and paving the way for a more agile computing environment. 
The second building block is the hierarchical resource continuum.
Upon establishing peering relationships, local clusters gain logical access to remote resource slices, exposed and available for application offloading.
This method accommodates the limited knowledge propagation and multi-ownership constraints inherent in a computing continuum environment.
The abstraction of peered clusters into virtual big nodes facilitates resource optimization, simplifying borrowing computational capacity, and permits application deployment on user-defined slices of the infrastructure.

\textit{Communication} transparency is another crucial aspect of the computing continuum.
Given the varying nature of microservice communication within Kubernetes clusters, the need for different primitives and explicit configurations may arise.
However, liquid computing mitigates this complexity by proposing a virtual cluster that spans across multiple real clusters, enabling seamless microservice interaction regardless of location.
This process decreases the need for intricate configurations, thus reducing potential errors.
In this context, the network continuum serves as the third essential building block.
In a liquid computing environment with applications spread across multiple clusters, this mechanism glues together separate network fabrics into a virtual network continuum, hence facilitating transparent communication between microservices, irrespective of their physical location.
The network fabric transparently handles potential configuration conflicts (e.g., overlapping IP addressing spaces), ensuring seamless and efficient communication across the liquid computing environment.

Liquid computing also ensures \textit{resource availability} transparency.
In contrast to traditional scenarios where microservices can only utilize resources within their respective clusters, liquid computing enables a service to access all resources within a virtual domain, irrespective of their physical location.
As such, a service can dynamically scale based on resource availability within the entire virtual infrastructure, eradicating traditional cluster boundaries.
The final two building blocks of this approach are resource and service reflection, and the storage and data continuum.
Resource and service reflection ensures that control plane information is present in both local and remote clusters, facilitating the execution of workloads.
On the other hand, the storage and data continuum addresses the needs of stateful workloads by providing persistent storage and data proximity, leveraging the concept of data gravity to minimize network traffic, reduce latency, and ensure regulatory compliance.
Together, these elements promote seamless workload execution and optimal data management across the liquid computing environment.

The outlined building blocks jointly establish a robust basis for actualizing liquid computing.
Although the primary focus herein is Kubernetes, the fundamental principles and strategies are universally applicable across varying orchestration platforms.
This adaptability facilitates the integration of liquid computing principles within a broad range of computing environments, thereby accommodating the distinct needs and traits of various deployment scenarios.

\section{Research challenges and Conclusion}

The journey towards the full realization of the computing continuum is well underway, as evidenced by the advent and ongoing development of liquid computing.
Embracing a dynamic, seamless, and transparent paradigm, the defined features embody the key characteristics that define the computing continuum.
By fostering deployment, communication, and resource availability transparency, it ensures a flexible, efficient, and integrated computing environment, weaving together previously isolated computing infrastructures.

Despite these significant strides, it is imperative to note that the actualization of a complete computing continuum is an ongoing endeavor.
Several challenges lie on the horizon that warrant further research.
Decentralization of control calls for innovative strategies to maintain stability and security in an ever-evolving, fluid landscape.
The efficacy of communication and resource availability transparency can be further bolstered by delving deeper into optimization strategies.

Moreover, the dynamic nature of data and services traversing across jurisdictions within a computing continuum environment amplifies the need for robust strategies to maintain data privacy and sovereignty.
In addition, we have to address the non-trivial problem of cross-domain authentication and authorization, mutual trust, and the run-time security at large (e.g., the hosting cluster should be protected from potential malicious actions carried out by the guest workloads, and the guest workload should be guaranteed from malicious actions, such as code tampering or stealing, potentially carried out by the hosting infrastructure).
Simultaneously, the handling of stateful workloads across multiple clusters within the storage and data continuum presents another rich area of exploration.

Furthermore, the reach of the computing continuum extends beyond Kubernetes. Consideration and integration with other orchestration platforms offer potential opportunities for enhancing the universality of the computing continuum, adding new layers of functionality and flexibility.

In essence, the computing continuum, while not fully actualized, is closer to reality than ever before, courtesy of the depicted principles and practices, and thanks to promising open-source projects such as Liqo.io \cite{liqo} and ongoing European projects fully committed to this vision, such as FLUIDOS \cite{fluidos}.
However, to proclaim the arrival of the complete computing continuum at this juncture would be premature.
The identified research challenges underscore the need for continued innovation and exploration in the field.
Through this persistent effort, we move steadily closer to the full realization of the computing continuum, a promising future where computing resources are seamlessly integrated, dynamic, and optimally utilized.

\section*{Acknoledgements}
This work was supported by European Union’s Horizon Europe research and innovation programme under grant agreement No 101070473, project FLUIDOS (Flexible, scaLable, secUre, and decentralIseD Operating System).

This research was conducted as part of Jacopo Marino Ph.D. programme, under the financing of the Piano Nazionale di Ripresa e Resilienza (PNRR) and the NextGenerationEU initiative. 

\bibliographystyle{IEEEtran.bst}
\bibliography{biblio}

\end{document}